\documentclass[twocolumn,prc,aps,showpacs]{revtex4}
\usepackage{graphicx}

 \def\gsim{\mathrel{\rlap{\lower4pt\hbox{\hskip1pt$\sim$}}
 \raise1pt\hbox{$>$}}}

 \newcommand\la{\langle}
 \newcommand\ra{\rangle}
 \newcommand\beq{\begin{equation}}
 
 \newcommand\eeq{\end{equation}}
 \newcommand\beqn{\begin{eqnarray}}
 \newcommand\eeqn{\end{eqnarray}}

\def\fm{\,\mbox{fm}}
\def\GeV{\,\mbox{GeV}}

\def\lsim{\mathrel{\rlap{\lower4pt\hbox{\hskip1pt$\sim$}}
    \raise1pt\hbox{$<$}}}         %less than or approx. symbol
\def\gsim{\mathrel{\rlap{\lower4pt\hbox{\hskip1pt$\sim$}}
    \raise1pt\hbox{$>$}}}         %greater than or approx. symbol

\def\fm{\,\mbox{fm}}
\def\GeV{\,\mbox{GeV}}

\def\s0{\sigma_0(s)}

\def\beq{\begin{equation}}
\def\eeq{\end{equation}}

\def\beqy{\begin{eqnarray}}
\def\eeqy{\end{eqnarray}}

\newcommand{\ber}{\begin{displaymath}}
\newcommand{\eer}{\end{displaymath}}
\newcommand{\bey}{\begin{eqnarray}}
\newcommand{\eey}{\end{eqnarray}}

\def\beq{\begin{equation}}
\def\eeq{\end{equation}}

\def\beqy{\begin{eqnarray}}
\def\eeqy{\end{eqnarray}}

\begin{document}
%\date{today}

\title{\bf Color transparency and suppression of high-\boldmath$p_T$ hadrons in nuclear collisions}

\vspace{1cm}

\author{B. Z. Kopeliovich}
\author{I. K. Potashnikova}
\author{Iv\'an Schmidt}
\affiliation{Departamento de F\'{\i}sica
Universidad T\'ecnica Federico Santa Mar\'{\i}a; and
\\
Instituto de Estudios Avanzados en Ciencias e Ingenier\'{\i}a; and\\
Centro Cient\'ifico-Tecnol\'ogico de Valpara\'iso;\\
Casilla 110-V, Valpara\'iso, Chile}
\begin{abstract}
\noindent 
The production length $l_p$ of a leading (large $z_h$) hadron produced in hadronization of a highly virtual high-$p_T$ parton is short because of the very intensive vacuum gluon radiation and dissipation of energy at the early stage of process.
Therefore, the main part of nuclear suppression of high-$p_T$ hadrons produced in heavy ion collisions is related to the survival probability of a colorless dipole propagating through a dense medium.
This is subject to color transparency, which leads to a steep rise with $p_T$ of the nuclear ratio  $R_{AA}(p_T)$, in good agreement with 
the recent data from the ALICE experiment at LHC, CERN. No adjustment, except the medium density is made, and the transport coefficient is found to be $\hat q_0=0.8\GeV^2/\fm$. This is close to the value extracted from the analysis of RHIC data for $J/\Psi$ suppression, but is an order of magnitude smaller than the value found from jet quenching data within the energy loss scenario.
Although the present calculations have a status of a postdiction, the mechanism and all formulas have been published, and are applied here with no modification, except the kinematics. At the same time, $p_T$-dependence of $R_{AA}$ at the energy of RHIC is rather flat due to the suppression factor steeply falling with rising $x_T$, related to the energy conservation constraints. This factor is irrelevant to the LHC data, since $x_T$ is much smaller.
\end{abstract}

%\date{\today}


\maketitle

\section{Introduction}

One of the first results of the heavy ion program at LHC is the observation by the ALICE experiment \cite{alice-data} of a strong nuclear suppression of high-$p_T$ charged hadrons.  
These data expose novel features compared with similar measurements at RHIC \cite{phenix,star}. First, the nuclear suppression factor $R_{AA}$ reaches significantly smaller values. This is not a surprise, since at the LHC energies hadrons originate mainly from hadronization of gluons, which have a larger color charge than quarks dominating at RHIC. Correspondingly, gluons dissipate energy with a higher rate. Second,
$R_{AA}(p_T)$ steeply rises with $p_T$, while it exposes a rather flat $p_T$-dependence in RHIC data.
The latter is affected by the restrictions imposed by energy conservation \cite{knpjs}. It was predicted \cite{paradigms} that
the production rate for hadrons and direct photons is suppressed in $pA$ and $AA$ collisions by the deficit of energy not only at forward rapidities, but also at large $x_T=2p_T/\sqrt{s}$. Here we concentrate on the interpretation of LHC data, which are free of these complications, since the values of $x_T$ are very small.

 The popular model explaining the observed suppression of $R_{AA}$ at high-$p_T$ relates it to the induced radiation energy loss by a parton propagating through the medium, which was created in the nuclear collision (e.g. see in \cite{miklos}). This energy-loss scenario is based on the unjustified assumption that hadronization of the parton lasts longer than the time of propagation through the medium, and that the detected hadron is always produced outside the medium. However, because of the steeply falling $p_T$ dependence of the cross section, most of high-$p_T$ hadrons carry a large fraction $z_h$ of the jet momentum, and 
energy conservation constraints 
the production length $l_p$ for such leading hadrons. It is expected to be rather short even within the simple string model \cite{k-nied,bg}, and should be much shorter in the case of intensive vacuum gluon radiation by a high-$p_T$ parton. 

One should clearly distinct between the production time scales for a colorless dipole (pre-hadron) and the final hadron.
The former signals on color neutralization, which stops the intensive energy loss caused by vacuum radiation following
the hard process, while the latter is a much longer time taken by the dipole to gain the certain hadronic mass, i.e. to develop the hadron wave function. While the former contracts $\propto(1-z_h)$ at large fractional momentum $z_h$ of the hadron, the latter keeps rising $\propto z_h$. These two time scales are frequently mixed up. The shortness of the production lengths at large $z_h$ is dictated by energy conservation. Indeed, a parton originated from a hard reaction intensively radiates losing energy, and this should cannot last long, otherwise the parton energy will drop below the energy of the detected hadron.

One should also distinct between the mean hadronization time of a jet, whose energy is shared between many hadrons, and specific events containing  a leading hadron with $z_h\to1$. Production of such a hadron in a jet is a small probability fluctuation, usually associated with large rapidity gap events. The space-time development of such an unusual jet is different from the usual averaged jet.

The controversy between the models with short and long production times has been under debate the last two decades (see in \cite{within,paradigms}), but no proof of a long time scale has 
been proposed so far, to the best of our knowledge. Of course, an experimental verification would be most convincing. Data on high-$p_T$ hadron production in heavy ion collisions provides a rather poor test of the models. Too many uncertainties are involved,
the medium properties are unknown, and their variation in space and time is based on simplified,even ad hoc models.
The fractional energy $z_h$ is not known, but enters the convolution of the initial parton distribution, hard cross section and the fragmentation function.
The important contribution of initial state effects (cold nuclear matter) can be only calculated within models. 

Probably the best way to study the space-time development of hadronization is inclusive hadron production in DIS on nuclei at large Bjorken $x$. In this case the medium density and its spacial distribution is well known. The fractional energy $z_h$ of the hadron is directly measured. A good model should predict the nuclear modification factor with no fitting.  Indeed, such a prediction was provided in \cite{knp} within the hadronization model with a finite production length. Later the first data from the HERMES experiment \cite{hermes} confirmed well this prediction. Also HERMES results on broadening of transverse momentum, sensitive to the production length \cite{within}, were explained well in \cite{broad-pir}.
On the other hand, the comprehensive study of nuclear effects within the pure energy loss scenario performed recently in \cite{wang} led to  striking disagreement with data for leading hadrons.

\begin{figure}[htb]
 \includegraphics[height=6.0cm]{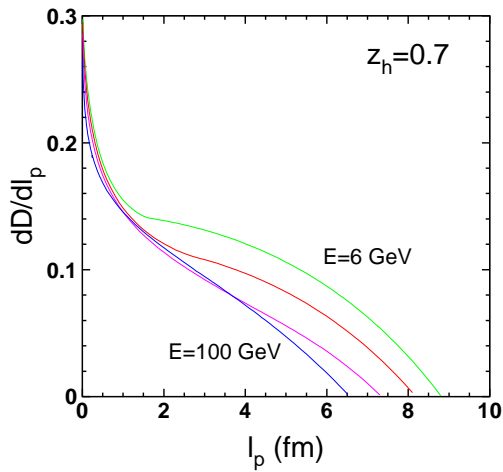}
\caption{ \label{lp-dep} 
The pre-hadron production length distribution $\partial D(z)/\partial l_p$
(in arbitrary units) for for a quark jet with energies $E=k_T=6,\ 10,\  20,\ 100\GeV$ (from top to bottom) and $z_h=0.7$. }
 \end{figure}

Here we rely on the model \cite{jet-lag} for the production time distribution of leading hadrons in a jet, 
produced at the mid rapidity. In this case the initial parton energy and virtuality are equal,
\beq
E=Q=k_T=\frac{p_T}{z_h} ,
\label{20}
\eeq
where $k_T$ and $p_T$ are the transverse momenta of the parton initiating the jet and of the detected hadron, respectively.
An example of the $l_p$ distribution at $z_h=0.7$ and different quark jet energies is shown in Fig.~\ref{lp-dep}.
One can see that the $l_p$-distribution narrows with energy, but is leveling off at high energies. 
This happens due to compensation of several effects \cite{jet-lag}, acting in opposite directions. The  Lorentz factor makes $l_p$ longer with energy, while the increasing virtuality  gives rise to a more intensive gluon radiation and energy loss in vacuum, leading to a shorter $l_p$. Moreover, the Sudakov suppression, essential at large $z_h$, also shortens $l_p$.

\begin{figure}[b]
 \includegraphics[height=5.5cm]{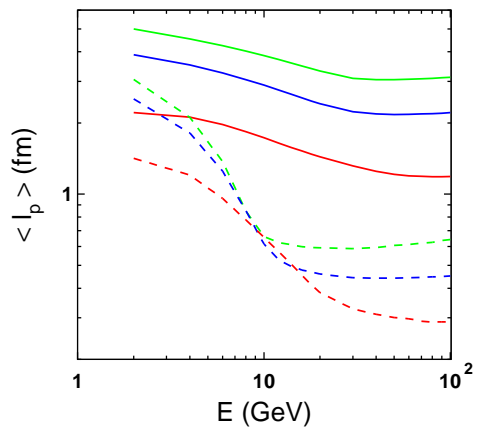}
\caption{ \label{mean-lp} 
The mean production length as function of energy for quark (solid curves) and gluon (dashed curves) jets.
In both cases the curves are calculated at $z_h=0.5,\ 0.7,\ 0.9$ (from top to bottom). }
 \end{figure}

The mean value $\la l_p\ra$ for quark jets, weighted with the distribution $dD/dl_p$, is depicted by solid curves in Fig.~\ref{mean-lp}, as function of energy, for $z_h=0.5,\ 0.7,\ 0.9$.
 Indeed, $\la l_p\ra$ saturates at high jet energies $E=k_T$. 
 
 Notice that in DIS the mean production length at fixed $Q^2$ rises linearly as function of energy. This case is quite different from high-$p_T$, where the virtuality increases with energy.
 
For gluon jets the energy loss is larger and the Sudakov suppression stronger due to the Casimir factor. This leads to a shorter production length. The mean value of $l_p$ for gluon jets as function of energy is depicted in Fig.~\ref{mean-lp} by dashed curves, for several values of $z_h$. We see that at high energies $\la l_p\ra <1\fm$. Thus, one can say that the pre-hadron is created almost instantaneously, because its production time is shorter than the expected time of medium creation, $t_0\sim 1\fm$.

An interesting possibility of a very dense medium was considered in \cite{call,paradigms}. If the medium were so dense that the mean free path of the produced dipole was vanishingly small, the nuclear suppression factor $R_{AA}$ would be proportional to $\la l_p^2\ra$, and could be predicted in a parameter free way.
Although such a possibility does not contradict RHIC data on jet quenching \cite{call,paradigms}, the recent study of $J/\Psi$ suppression in heavy ion collisions \cite{psi,psi-bnl} found that the density of the medium is rather low. 
The maximal value of transport coefficient reached at the time scale of the medium creation $t_0\sim1\fm$ was found to be quite low, $\hat q_0\sim 0.2\GeV^2/\fm$, an order of magnitude smaller than follows from hadron suppression data interpreted within the energy loss formalism \cite{e-loss}.
Thus, the scenario of a very dense medium is not supported by data, and one should study propagation of colorless dipole through a moderately opaque medium.

Here we employ the same description of dipoles in a medium, as was used for $J/\Psi$ suppression \cite{psi}, and apply it to high-$p_T$ processes. The main difference is the time scale of formation of the final hadron: while it is very short for a heavy and slow $J/\Psi$ \cite{psi,psi-bnl}, the formation of the wave function of a light $\bar qq$ dipole
moving with a high momentum is rather long. Correcting for this, we apply the same description to the suppression nuclear factor $R_{AA}(p_T)$ for hadrons produced in central lead-lead collisions at LHC, and achieve good agreement with the recent data from the ALICE experiment \cite{alice-data}. Moreover,
we found a value of the transport coefficient similar to what was found from $J/\Psi$ data at RHIC.

\section{Attenuation of a small size dipole in a medium}

A parton, which experienced a hard scattering with transverse momentum $k_T$, shakes off its color field
in the form of a forward cone of gluon radiation, and starts propagating along a new direction, lacking the soft part of its field up to transverse frequencies $\lsim k_T$. The parton starts regenerating its color field, with transverse size $r(l)$ expanding as function of path length $l$, starting from the initial small size $r_0\sim1/k_T$ at $l=0$.  At some distance $l_p$
a colorless dipole, pre-hadron, is produced with a size of the order of the  
transverse size $r(l_p)$ of the regenerated field. We describe the regeneration process and gluon radiation within the dipole approach \cite{kst1,kst2}.

A small size dipole is expanding so fast that its initial size is
quickly forgotten. Indeed, the speed of expansion of a dipole
correlates with its size: the smaller the dipole is, the faster it
is evolving. This is controlled by the uncertainty principle, $q\sim
1/r$.
\beq \frac{dr}{dt}=2v_T=\frac{2q}{E}\approx
\frac{2}{E\,r}, 
\label{30}
\eeq
where $E=p_T$ is
the dipole energy in the c.m. of the collision; $v_T$ and $q\sim 1/r$ are the transverse velocity and   momentum of the quark relative to the dipole momentum direction.
The solution of this equation reads \cite{psi,psi-bnl}, 
\beq
r^2(t)=\frac{4\,t}{p_T} + r_0^2, 
\label{40} 
\eeq 
where
$r_0\sim 1/p_T$ is the initial dipole size,
neglected in what follows.

Propagation of a dipole over path length $L$ in a medium is characterized by a
survival probability
\beq 
S(L)=\exp\left[-\int\limits_0^L dl\,
\sigma[r(l)]\,\rho(l)\right],
\label{60}
\eeq
where the dipole cross section $\sigma(r)$ times the medium density $\rho$ is the attenuation rate of the dipole.

The dipole cross section for small dipoles is $\sigma(r)=C\,r^2$.
 The factor $C$, for
dipole-proton interactions, is known from DIS data. Its value for a
hot medium is unknown, as well as the medium properties. 
It is convenient to express it in terms of the so called transport coefficient, which is the broadening of a parton in the medium over the path length $1\fm$.
Indeed, the same factor C controls both the dipole cross section and broadening of a quark propagating through
the medium \cite{jkt,dhk}.
So, the factor $C$ in the dipole cross section is related to the transport
coefficient $\hat q$ \cite{bdmps}, which is the in-medium broadening per
unit of length,
\beq 
C=\frac{\hat q}{2\,\rho}. 
\label{90} 
\eeq
Then, using (\ref{40}) (with $t=l$) one can represent the survival probability of the dipole in the medium,
Eq.~(\ref{60}), as
\beqn
S(L)&=&\exp\left[- {1\over2}
\int\limits_0^L dl\, \hat q(l)\,r^2(l)\right]
\nonumber\\ &=&
\exp\left[- {2\over p_T}
\int\limits_0^L dl\, l\,\hat q(l)\right].  
\label{100} 
\eeqn

Now we are in a position to calculate the nuclear attenuation factor for a high $p_T$ hadron produced in heavy ion collision. For central ($b=0$) collisions of two identical nuclei one should integrate over the impact parameter $\vec\tau$ of the hard collision, with a weight factor $T_A^2(\tau)$, where $T_A(\tau)=\int_{-\infty}^{\infty}dz\,\rho_A(\vec\tau,z)$ is the nuclear thickness function, the integral of nuclear density along the collision direction:
\beq
R_{AA}(b=0,p_T)=\frac{\int\limits_0^\infty d^2\tau\,T_A^2(\tau)\,R_{AA}(\tau,p_T)}
{\int\limits_0^\infty d^2\tau\,T_A^2(\tau)}.
\label{500}
\eeq
The factor $R_{AA}(\tau,p_T)$ is the nuclear suppression factor corresponding to production of a high-$k_T$ parton
at impact parameter $\vec\tau$, propagating then over a path length $\la l_p\ra$, radiating gluons and losing energy, and eventually producing a colorless dipole pre-hadron with transverse momentum $\vec p_T=\vec k_T/z_h$, which
propagates through the nucleus evolving its size according to Eq.~(\ref{40}). We rely on the above evaluation of $\la l_p\ra$ in vacuum, since the medium-induced energy loss is much smaller than the vacuum one. Besides, induced energy loss can make $l_p$ only shorter, which will not affect the further calculations.
This suppression factor has the form \cite{psi,psi-bnl},
\beq 
R_{AA}(\vec\tau,p_T)\Bigr|_{b=0}= 
\int\limits_0^\pi
\frac{d\phi}{\pi} \exp\Biggl[-\frac{2}{p_T}
\int\limits_{l_{max}}^\infty dl \,l\,\hat q(\vec\tau+\vec l)
\Biggr],
\label{140} 
\eeq
where $l_{max}=max\{l_p,l_0\}$. Here $t_0=l_0\sim1\fm$ is the time scale of creation of the medium resulted from gluon radiation at mid rapidities in heavy ion collisions. Since the production length for a gluon jet is short, $\la l_p\ra\lsim l_0$, its actual value is not important.

The medium density is time dependent, and
is assumed to dilute as  $\rho(t)=\rho_0\,t_0/t$ due to the
longitudinal expansion of the produced medium. 
Correspondingly, the transport coefficient depends on impact parameter and time (path length) as \cite{frankfurt},
\beq 
\hat q(l,\vec b,\vec\tau)=\frac{\hat
q_0\,l_0}{l}\, \frac{n_{part}(\vec b,\vec\tau)}{n_{part}(0,0)},
\label{120} 
\eeq
where $n_{part}(\vec b,\vec\tau)$ is the number of participants;
$\hat q_0$ corresponds to the
maximum medium density produced at impact parameter $\tau=0$ in
central collisions ($b=0$) at the time $t=t_0=l_0$ after
the collision. In what follows we treat the transport coefficient
$\hat q_0$ as an adjusted parameter.

\section{Results vs data}

\begin{figure}[htb]
 \includegraphics[height=7cm]{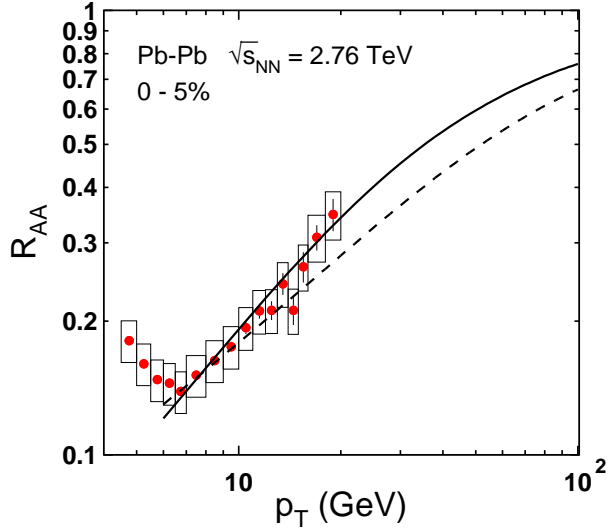}
\caption{\label{pt-dep} ALICE data for central, 0-5\%, lead-lead collisions at $\sqrt{s}=2.76\GeV$
\cite{alice-data}. The solid curve corresponds to Eqs.~(\ref{140})-(\ref{500}) calculated with $l_p\leq l_0=1\fm$ and $\hat q_0=0.8\GeV^2/\fm$. The dashed  curve, calculated with $l_p=2\fm$, demonstrates sensitivity to $l_p$.
}
 \end{figure}

The results of the calculation with Eqs.~(\ref{500})-(\ref{140}) with $\hat q_0=0.8\GeV^2/\fm$ are shown by the solid curve in Fig.~\ref{pt-dep}, in comparison with ALICE data \cite{alice-data}.
Except for $\hat q_0$, no further adjustment was done, and agreement with data at large $p_T>7\GeV$ is pretty good. 
Moreover, the transport coefficient turns out to be of the same order as was found in \cite{psi} from data on $J/\Psi$ production.

The dashed curve in Fig.~\ref{pt-dep} demonstrates the sensitivity to $l_p$. It is calculated at $\la l_p\ra=2\fm$ and accordingly adjusted $\hat q_0=1.6\GeV^2/\fm$.
Since at large $p_T\gsim100\GeV$ valence quarks with larger $l_p$ should dominate, the rise of $R_{AA}$ with
$p_T$ should slightly slow down deviating from the solid curve towards the dashed one.

We do not attempt here to describe $R_{AA}$ at $p_T<7\GeV$, since the dynamics becomes much more complicated.
First, the production length becomes several time longer, as is depicted in Fig.~\ref{mean-lp}. Second, one should take into account the Cronin effect in $AA$ collisions, which is poorly known because of the large fraction of baryons 
in detected charged hadrons.

For the same reason, application of this mechanism to high-$p_T$ hadron production at RHIC goes beyond the scope of this paper, since it involves a more complicated and model dependent dynamics. Indeed, it was demonstrated in \cite{knpjs} that for particle production at forward rapidities, $x_F\gsim 0.1$, energy conservation becomes an issue.
It causes additional nuclear suppression, steeply increasing with $x_F$. The same, of course, should happen at large $x_T=2p_T/\sqrt{s}$. It was demonstrated in \cite{paradigms} that this mechanism leads to suppression of high-$p_T$ hadrons and direct photons in $pA$ and $AA$ collisions. Therefore, besides other reasons (different kinematics, valence quark dominance, etc.), this mechanism makes the $p_T$ dependence of $R_{AA}(p_T)$ significantly flatter at RHIC than at LHC.  Detailed calculations for the RHIC energy domain will be published elsewhere.

\section{Summary}

Although these calculations have a status of post-diction, the mechanism and all formulas have been already published, and we apply them here with no specific modification, except for the kinematics. The dynamics of nuclear suppression of high-$p_T$ hadrons produced in central lead-lead collisions at $\sqrt{s}=2.76\GeV$ is based on the shortness of the production length of a pre-hadron, and its development and propagation through a dense medium.
We performed calculations within the same scheme as was used for the analysis \cite{psi} of $J/\Psi$ production data. Moreover, we arrived at a value of the transport coefficient, characterizing the properties of the medium, which is
pretty close to the value extracted from $J/\Psi$ data, and is an order of magnitude smaller than what has been obtained from analyses of jet quenching RHIC data based on the energy loss scenario.

These calculations can be improved by replacing the simplified description of the dipole evolution, Eq.~(\ref{40}), with the rigorous quantum-mechanical approach \cite{kz91} based on the path integral technique.
This description can be also applied to RHIC data on jet quenching, but one should introduce a model dependent suppression factor, which is related to the constraints on nuclear parton distributions imposed by energy conservation
\cite{knpjs}. These further developments of the present approach will be published elsewhere.

\begin{acknowledgments}

We are grateful to Jan Nemchik for pointing out a dimension related typo in the code. We are also 
thankful to Martin Poghosyan who provided us with the tables of ALICE data. This work was supported in part
by Fondecyt (Chile) grants 1090236, 1090291 and 1100287, by DFG
(Germany) grant PI182/3-1, and by Conicyt-DFG grant No. 084-2009.

\end{acknowledgments}

\end{document}